\documentclass[12pt,preprint]{elsarticle}
\usepackage[latin9]{inputenc}
\usepackage{float}
\usepackage{amstext}
\usepackage{amssymb}
\usepackage{graphicx}
\usepackage{esint}

\makeatletter

\providecommand{\tabularnewline}{\\}

\usepackage{url}

\newcounter{bla}

\journal{Computer Physics Communications}

\makeatother

\begin{document}
\begin{frontmatter}



\title{ZMCintegral-v5: Support for Integrations with the Scanning of Large
Parameter Grids on Multi-GPUs}


\author[a]{Jun-Jie Zhang\corref{author}}

\author[a]{Hong-Zhong Wu\corref{author}}

\cortext[author]{Both authors contributed equally to this manuscript.}

\address[a]{Department of Modern Physics, University of Science and Technology
of China}


\begin{abstract}
In this updated vesion of ZMCintegral, we have added the functionality
of integrations with parameter scan on distributed Graphics Processing
Units(GPUs). Given a large parameter grid (up to $10^{10}$ parameter
points to be scanned), the code will evaluate integrations for each
parameter grid value. To ensure the evaluation speed, this new functionality
employs a direct Monte Carlo method for the integraion. The Python
API is kept the same as the previous ones and users have a full flexibility
to define their own integrands. The performance of this new functionality
is tested for both one node and multi-nodes conditions.
\end{abstract}

\end{frontmatter}



\noindent \textbf{PROGRAM SUMMARY} 

\noindent \begin{small} {\em Manuscript Title:} ZMCintegral-v5:
Support for Integrations with the Scanning of Large Parameter Grids
on Multi-GPUs \\
 {\em Authors:} Jun-Jie Zhang;Hong-Zhong Wu \\
 {\em Program Title: ZMCintegral} \\
 {\em Journal Reference:} \\
{\em Catalogue identifier:} \\
{\em Licensing provisions:} Apache License Version, 2.0(Apache-2.0)\\
{\em Programming language:} Python \\
{\em Operating system:} Linux 
{\em Classification:} 4.12 Other Numerical Methods \\
{\em External routines/libraries:} Numba; Ray; \\
{\em Nature of problem:} Easy to use python package for integrations
with large parameter grids using Monte Carlo method on distributed
GPU clusters.\\
{\em Solution method:} Direct Monte Carlo method and distributed
computing.\\

\section{Introduction}

In the real application of ZMCintegral, we find that for many cases
users usually have integrations with various parameters. The previous
versions\cite{Wu2019} of our package mainly fouces on high dimensional
integration, and lack the performance for a search of the parameter
grids. Furthermore, there are many ereas that require the integration
involving parameters while the dimensionality is not very high. For
example, the solving of GAP equations in finite temperature field\cite{Shi2014,Shi2016},
the branching fractions predictions in meson decay\cite{Cui2019},
the solving of transport equations in phase space for quark gluon
plasma\cite{Kurkela2019,DeGroot:1980dk,Xu2005}, the calculation of
global polarization at different coherent length in heavy iron collisions\cite{Zhang2019},
etc. Therefore, we add this new functionality of parameter grid search
to version-5. In fact, there have been many packages\cite{Back2018,James:1994vla,GPLepage1978,Jadach2003}
(for specific use), and also commercial softwares (Mathematica, Matlab,
etc.) that have good performance for parameter grid search with CPU
devices. Also, with the development of GPU CUDA\cite{cuda01}, integrations
on CUDA for decoupled ODEs (ordinary differential equations) with
various initial conditions (can be seen as parameters), has also been
developed\cite{Niemeyer2014}.

For our scheme, we mainly work on a general-use Python package with
Monte Carlo integration on multi-GPUs of distributted clusters. In
this updated version, we have added the functionality where users
can provide a large series of parameters for the integration, and
the package requires Numba\cite{Kwan2015} and Ray\cite{PMoritz2018}
to be pre-installed. Since this functionality is mainly for the search
of the parameter grid, we have merely applied the direct (or simple)
Monte Carlo method\cite{mcbook} without stratified sampling and heuristic
tree search to ensure the speed performance. The source codes and
manual can be found in Ref. \cite{zmc_github}. The Python API is
kept the same as the previous ones and users have a full flexibility
to define their own integrands.

In this paper, we first introduce the general structure of the functionality
of this new version. Then, few examples have been introduced to demonstrate
the performance of the speed and accuracy. For users who have a large
parameter grid (up to $10^{10}$ parameter grid points), this new
functionality (version-5) will be suitable.

\section{\label{sec:Integration-with-parameter}Integration with parameter
grid}

In our code, the general form of the integration with parameters is
defined as
\begin{eqnarray}
f(\mathbf{x}) & = & \int\text{\ensuremath{d\mathbf{y}}}g(\mathbf{y};\mathbf{x}),\label{eq:general form}
\end{eqnarray}
where vector $\mathbf{y}=(y_{1},\cdots y_{N})$ is the integration
variable, $\int\text{\ensuremath{d\mathbf{y}}}\equiv\int\prod_{k=1}^{N}dy_{k}$,
and vector $\mathbf{x}=(x_{1},\cdots x_{M})$ is the grid point of
the parameter grid. Each parameter $x_{i}$ in $\mathbf{x}=(x_{1},\cdots x_{M})$
takes values from a list, which contains the values that need to be
scanned for this specific parameter $x_{i}$.

With any given $\mathbf{x}$, the integration will be evaluated using
the direct Monte Carlo method. Currently, it is time consuming to
perform a stratified sampling with large parameter grids. Hence, to
increase speed, we only consider the application of a direct MC, and
limit the usage for this version to large parameter grid and small
dimensional integration . For higher dimensional integration with
small parameter grid , the previous versions\cite{Wu2019} can take
the role.

The same as the previous versions, we haved used the package Ray\cite{PMoritz2018}
and Numba\cite{Kwan2015} to perform multi-GPU calculations on distributed
clusters. The returned result is a multi-dimensional grid with each
element being the integrated values.

In the real calculation, we cut the parameter grid into several batches.
Each batch is fed into one GPU device for evaluation. A task can either
be assigned sequentially to one GPU or parallelly to several distributed
GPUs. The integration is performed on every GPU thread, hence, each
thread gives the integration value of one parameter. This porcess
is shown in Fig. \ref{fig:Schematic-of-distributed}.

\begin{figure}

\begin{centering}
\includegraphics[scale=0.5]{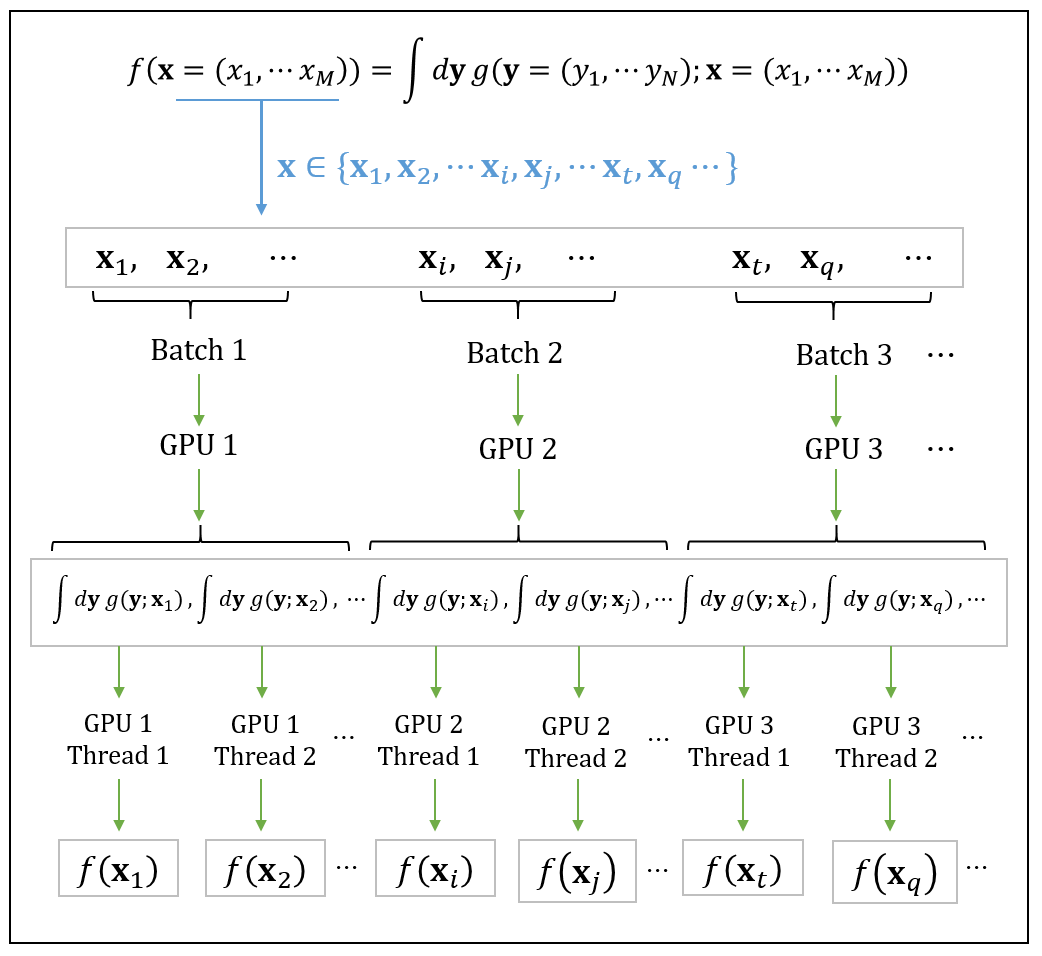}
\par\end{centering}
\caption{\label{fig:Schematic-of-distributed}Schematic diagram of distributed
GPU evaluation. Given the parameter grid value $\mathbf{x}_{1},\mathbf{x}_{2}\cdots$,
the corresponding integrations $\int d\mathbf{y}g(\mathbf{y};\mathbf{x}_{1})$,
$\int d\mathbf{y}g(\mathbf{y};\mathbf{x}_{1})\cdots$ can be evaluated
in different GPU threads via direct Monte Carlo method. In our algorithm,
each batch is assigned to one GPU and each integration is assigned
to one thread.}

\end{figure}

\section{Results and performance}

\subsection{\label{subsec:Test-on-one}Test on one node}

The hardware condition for this node is Intel(R) Xeon(R) Silver 4110
CPU@2.10GHz CPU with 10 processors + 1 Nvidia Tesla V100 GPU.

We test our code with an oscillating integrand
\begin{eqnarray}
f(x_{1},x_{2},\cdots x_{M}) & = & (\prod_{k=1}^{N}\int_{0}^{10}dy_{k})\text{sin}(\sum_{j}^{N}y_{j}+\sum_{l=1}^{M}x_{l}),\label{eq:one node test}
\end{eqnarray}
where $N$ is the dimension of the integrations. Each parameter $x_{i}$
in $\mathbf{x}=(x_{1},\cdots x_{M})$ takes values from the list $\{0,1,2,\cdots99\}$,
which contains the values that need to be scanned for this specific
parameter $x_{i}$. Therefore, the parameter grid has $100^{M}$ points.
Hence we need to scan over $100^{M}$ points, with each point containing
an integration of $N$ dimensions.

Here we choose $M\in\{1,2,3,4\}$ and $N\in\{1,2,3,4\}$to see the
time consumption of HToD(host to device), DToH(device to host) and
the total evaluation. Since the theoretical value of this integration
can be obtained directly, we also compare our results with the theoretical
ones. We introduce the relative error $\delta r$ as
\begin{eqnarray}
\delta r & \equiv & \frac{1}{100^{M}}\sum_{i=1}^{100^{M}}\frac{|f_{\text{theoretical}}^{i}-f_{\text{ZMC}}^{i}|}{f_{\text{theoretical}}^{i}}.\label{eq:delta r}
\end{eqnarray}

Tab. \ref{tab:Performance-for-oscillating} and Fig. \ref{fig:Performance-for-oscillating}
show the results of the performance of our code on one node. We can
see from the upper-left part of Fig. \ref{fig:Performance-for-oscillating}
that for a large parameter scan, the dimension of the integration
should not be very large. In our case, we have chosen an oscillating
function and used merely $10^{4}$ sample points for each integration.
Therefore, we obtain a large error for dimension $N=4$. However,
in real ceses one can use more sample points (e.g. $\geq10^{4}$)
for higher dimensinal integrations (e.g. $N\geq4$, see Sec. \ref{subsec:Test-on-more}).
Since the parameter grid has $100^{M}$ points, with the increase
of $M$, the parameter grid size increases exponentially. This exponential
increase explains the close gap between $M=1,2,3$, and a large gap
between $M=4$ and $M=1,2,3$ in Fig. \ref{fig:Performance-for-oscillating}.
\begin{center}
\begin{table}[H]
\centering{}\caption{\label{tab:Performance-for-oscillating}Performance for oscillating
integrands on one node. This node contains one V100 GPU. $M\in\{1,2,3,4\}$
and $N\in\{1,2,3,4\}$ stand for the number of parameters and the
dimension of the integration respectively. The integration is performed
via the direct Monte Carlo method with 10000 samples. Number of batches
is set to 1. The data is plotted in Fig. \ref{fig:Performance-for-oscillating}.}
{\scriptsize{}}%
\begin{tabular}{ccccccccc}
\hline
{\scriptsize{}$(M,N)$} & {\scriptsize{}$(1,1)$} & {\scriptsize{}$(1,2)$} & {\scriptsize{}$(1,3)$} & {\scriptsize{}$(1,4)$} & {\scriptsize{}$(2,1)$} & {\scriptsize{}$(2,2)$} & {\scriptsize{}$(2,3)$} & {\scriptsize{}$(2,4)$}\tabularnewline
\hline
\hline
{\scriptsize{}$\delta r$} & {\scriptsize{}0.02512} & {\scriptsize{}0.25034} & {\scriptsize{}1.19949} & {\scriptsize{}6.19901} & {\scriptsize{}0.03379} & {\scriptsize{}0.16417} & {\scriptsize{}0.87280} & {\scriptsize{}4.12216}\tabularnewline
{\scriptsize{}HToD (ms)} & {\scriptsize{}0.72227} & {\scriptsize{}0.84085} & {\scriptsize{}0.70653} & {\scriptsize{}0.96030} & {\scriptsize{}1.14849} & {\scriptsize{}1.01440} & {\scriptsize{}1.19109} & {\scriptsize{}1.13764}\tabularnewline
{\scriptsize{}DToH (s)} & {\scriptsize{}0.00010} & {\scriptsize{}0.00011} & {\scriptsize{}0.00010} & {\scriptsize{}0.00010} & {\scriptsize{}0.00012} & {\scriptsize{}0.00013} & {\scriptsize{}0.00012} & {\scriptsize{}0.00013}\tabularnewline
{\scriptsize{}total time (s)} & {\scriptsize{}0.59343} & {\scriptsize{}0.57227} & {\scriptsize{}0.59330} & {\scriptsize{}0.60408} & {\scriptsize{}0.61858} & {\scriptsize{}0.59581} & {\scriptsize{}0.61055} & {\scriptsize{}0.62169}\tabularnewline
\hline
{\scriptsize{}$(M,N)$} & {\scriptsize{}$(3,1)$} & {\scriptsize{}$(3,2)$} & {\scriptsize{}$(3,3)$} & {\scriptsize{}$(3,4)$} & {\scriptsize{}$(4,1)$} & {\scriptsize{}$(4,2)$} & {\scriptsize{}$(4,3)$} & {\scriptsize{}$(4,4)$}\tabularnewline
\hline
\hline
{\scriptsize{}$\delta r$} & {\scriptsize{}0.02927} & {\scriptsize{}0.15052} & {\scriptsize{}0.78052} & {\scriptsize{}4.05029} & {\scriptsize{}0.08806} & {\scriptsize{}0.55222} & {\scriptsize{}3.55857} & {\scriptsize{}22.62904}\tabularnewline
{\scriptsize{}HToD (ms)} & {\scriptsize{}1.23856} & {\scriptsize{}1.19336} & {\scriptsize{}1.37538} & {\scriptsize{}1.41807} & {\scriptsize{}2.79922} & {\scriptsize{}2.68347} & {\scriptsize{}2.77383} & {\scriptsize{}2.64370}\tabularnewline
{\scriptsize{}DToH (s)} & {\scriptsize{}0.00200} & {\scriptsize{}0.00303} & {\scriptsize{}0.00197} & {\scriptsize{}0.00268} & {\scriptsize{}0.60880} & {\scriptsize{}0.61748} & {\scriptsize{}0.59681} & {\scriptsize{}0.59422}\tabularnewline
{\scriptsize{}total time (s)} & {\scriptsize{}1.26269} & {\scriptsize{}1.51221} & {\scriptsize{}1.78097} & {\scriptsize{}2.07094} & {\scriptsize{}64.8745} & {\scriptsize{}91.5998} & {\scriptsize{}118.275} & {\scriptsize{}147.347}\tabularnewline
\hline
\end{tabular}
\end{table}
\par\end{center}

\begin{center}
\begin{figure}[H]
\begin{centering}
\includegraphics[scale=0.25]{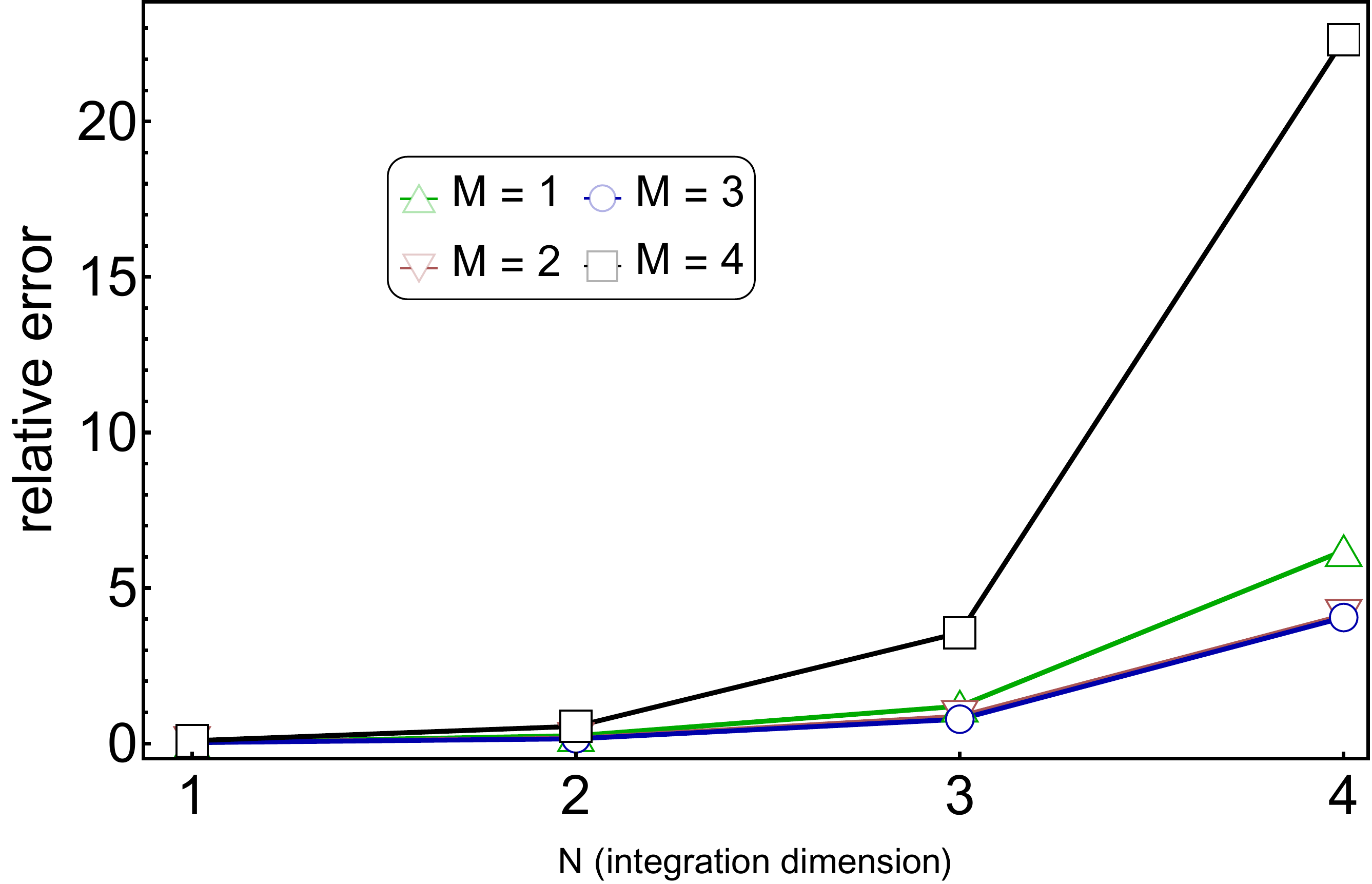}\includegraphics[scale=0.25]{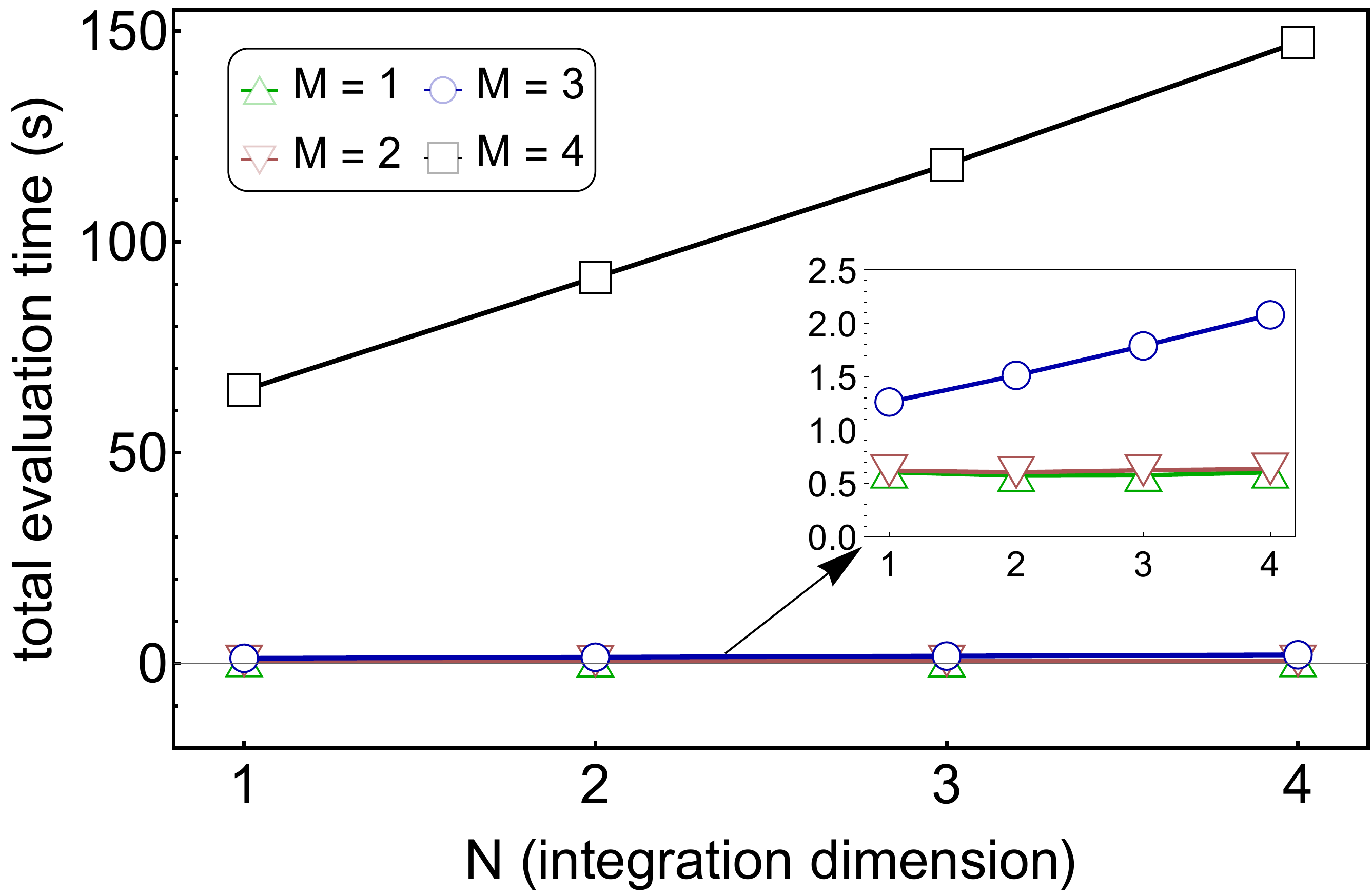}
\par\end{centering}
\centering{}\includegraphics[scale=0.25]{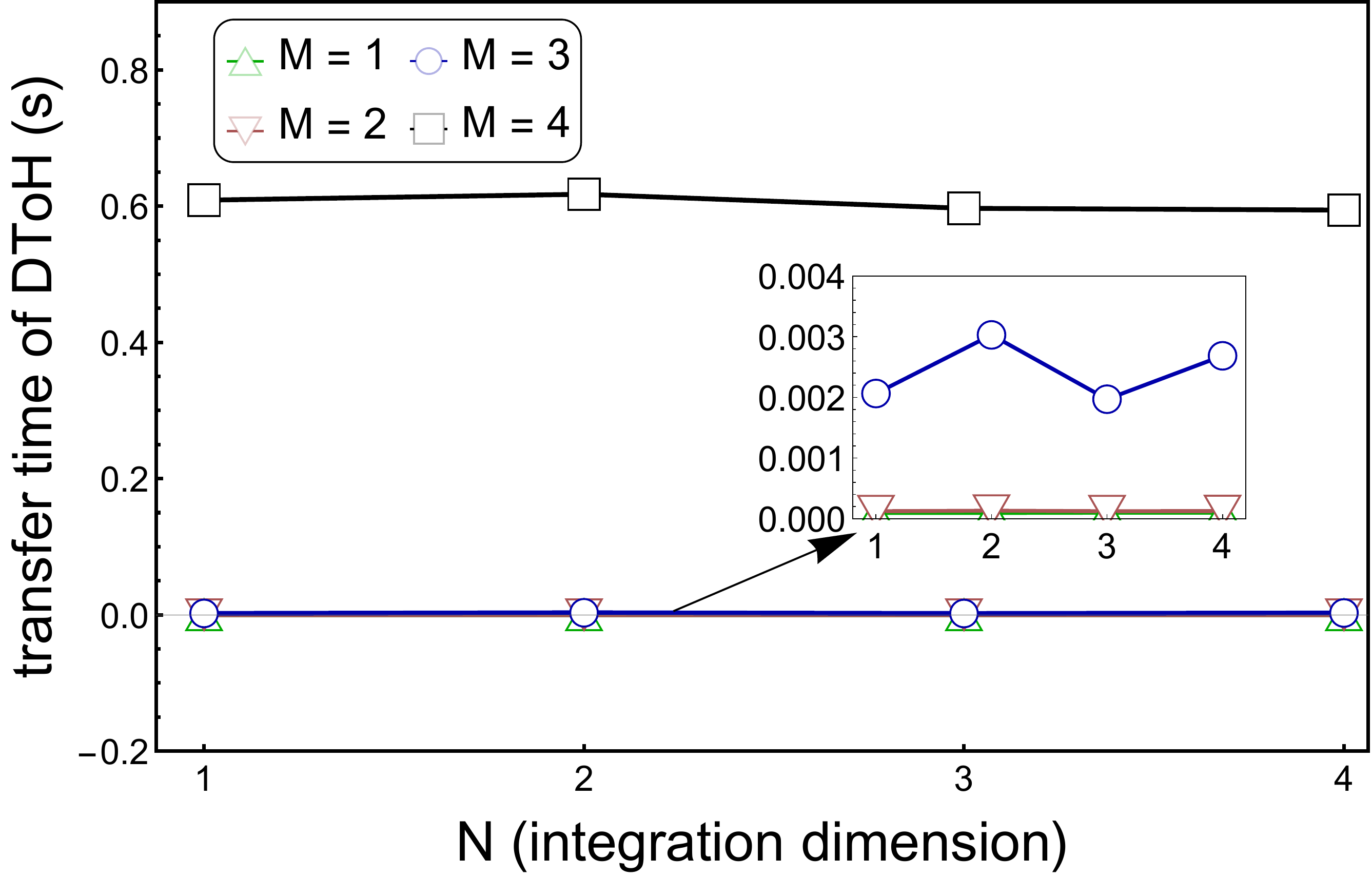}\includegraphics[scale=0.25]{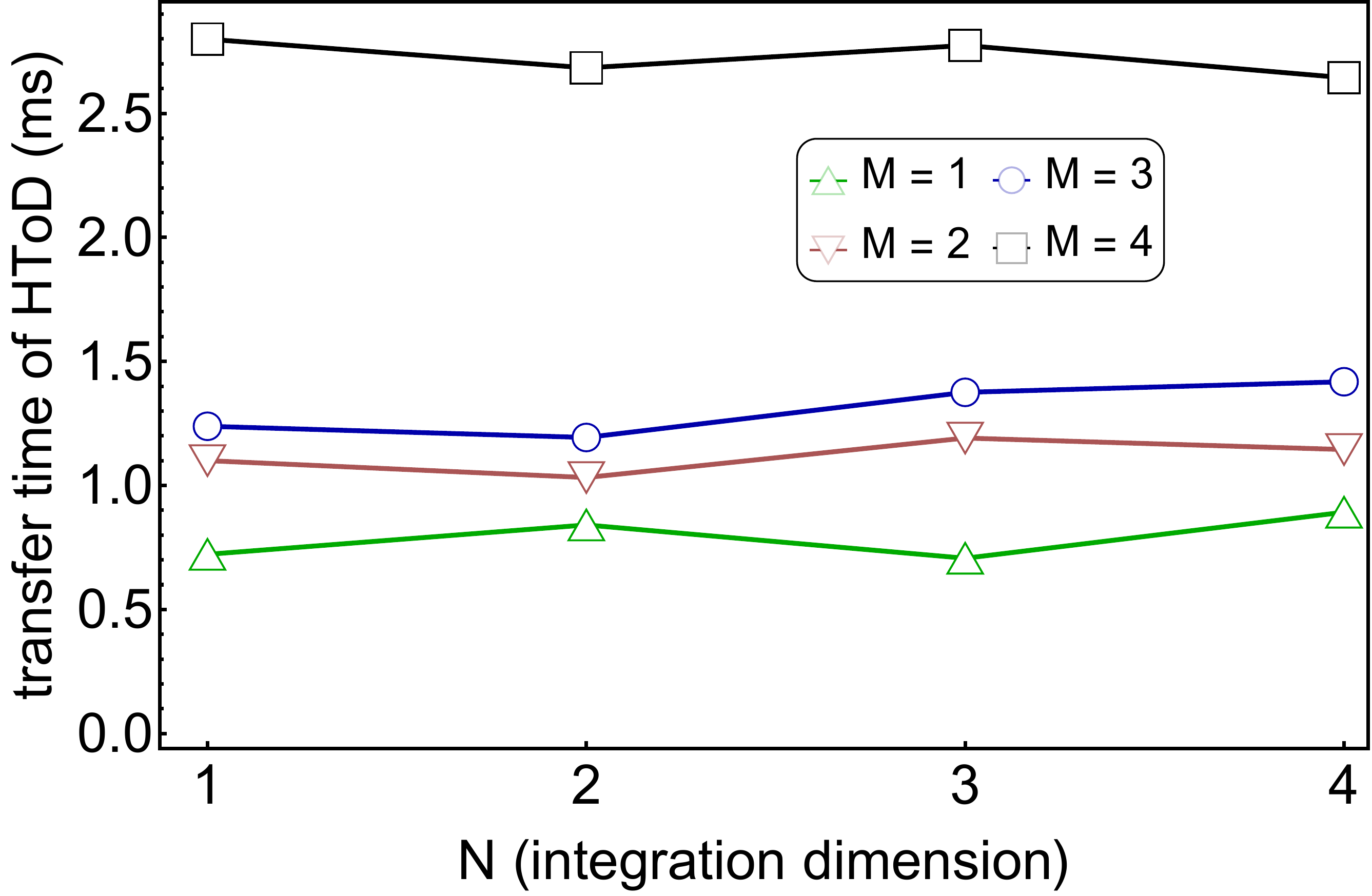}\caption{\label{fig:Performance-for-oscillating}Performance for oscillating
integrands on one node. This node contains one Nvidia Tesla V100 GPU.
The upper-left panel demonstrates the calculation error (defined in
Eq. (\ref{eq:delta r})) in terms of $M$ and $N$. It can be seen
that with the increase of the dimensions, the relative error increases
really fast. The upper-right panel shows the total evaluation time
of Eq. (\ref{eq:one node test}). We can see that the total time increases
with the increase of both $M$ and $N$. The lower panel dimenstrates
the transfer time of DToH (device to host) and HToD (host to device).
This transfer time mainly depends on the size of the parameter grid,
i.e. $M$, and is not sensitive to the integration dimension $N$.}
\end{figure}
\par\end{center}

\subsection{\label{subsec:Test-on-more}Test on multipole nodes}

We use totally three nodes in this section. The hardware condition
for the three nodes are Intel(R) Xeon(R) CPU E5-2620 v3@2.40GHz CPU
with 24 processors + 4 Nvidia Tesla K40m GPUs, Intel(R) Xeon(R) CPU
E5-2680 V4@2.40GHz CPU with 10 processors + 2 Nvidia Tesla K80 GPUs,
and Intel(R) Xeon(R) Silver 4110 CPU@2.10GHz CPU with 10 processors
+ 1 Nvidia Tesla V100 GPU. The K80 card can be seen as the combination
of two K40 cards in physical structure. These three nodes are in a
local area network.

Different from integrating one single function, where most computational
resources can be used to generate sample points, the parameter grid
search deals with millions of integrands of different parameters.
Therefore, the computational resources are used to loop thorugh the
parameters. As is introduced in Sec. \ref{sec:Integration-with-parameter},
a single thread needs to generate all sample points to perform the
integration of a certain parameter point. This means that the sample
points for each thread can not be very large. For this parameter scan
functionality, we suggest a number of sample points not exceed $10^{6}$
(for grid size \textasciitilde{} $10^{8}$) for one Tesla V100 (this
configuration will take roughly 6 hours with Intel(R) Xeon(R) Silver
4110 CPU@2.10GHz CPU with 10 processors + 1 Nvidia Tesla V100 GPU).
However, for users with large GPU clusters, this number can be set
higher. Since we adopt the direct Monte Carlo method to implement
the integration, the accuracy of the integration only depends on the
number of sample points. In the real application, users need to firstly
determine the number of sample points based on the desired accuracy.
Then start with some small value for the sample number, and increase
this value to see if the evaluation time is acceptable.

Now we test the performance of the code on three nodes with the integrand
\begin{eqnarray}
f(x_{1},x_{2},x_{3},x_{4}) & = & (\prod_{k=1}^{6}\int_{0}^{1}dy_{k})\text{sin}(\sum_{j}^{6}y_{j}+\sum_{l=1}^{4}x_{l}),\label{eq:three nodes test}
\end{eqnarray}
which is similar as Eq. (\ref{eq:one node test}), but with different
integration domain. We set $M=4$, $N=6$ and each parameter $x_{l}\in\{x_{1},x_{2},x_{3},x_{4}\}$
takes values in list $\{0,1,2,\cdots99\}$. To gain a rather stable
result of this sine function with domain $[0,10]^{6}$, we need $10^{10}$
sample points. $10^{10}$ ponits will take us a few years for a parameter
grid of size $10^{8}$ in the current cluster. Therefore, we choose
a domain $[0,1]^{6}$ where the function is not oscillating and limiting
the number of sample points to $10^{5}$. We emphasize that it is
still challenging to handle both the oscillation and large parameter
grid under the current GPU device.

In this test, we used all three nodes and perform the integration
of Eq. (\ref{eq:three nodes test}) independently for 10 times. The
results are shown in Tab. \ref{tab:Performance-three nodes}. It can
be seen that compared with the GPU evaluation time, HToD and DToH
are almost negligible. Meanwhile, the time consumption for task (data)
allocation and retrieve is negligible compared with the total evaluation
time. Therefore, most of the time are spent on GPU calculation and
the data transfer time is tiny in our test.

Furthermore, we also plot the relative error $\delta r$ in terms
of grid parameters $x_{1}$ and $x_{2}$ with
\begin{eqnarray}
\delta r(x_{1},x_{2}) & \equiv & \frac{1}{100^{2}}\sum_{x_{3},x_{4}}\frac{|f_{\text{theoretical}}(x_{1},x_{2},x_{3},x_{4})-f_{\text{ZMC}}(x_{1},x_{2},x_{3},x_{4})|}{f_{\text{theoretical}}(x_{1},x_{2},x_{3},x_{4})}.\label{eq:delta r partial sum}
\end{eqnarray}
It can be seen from Fig. \ref{fig:Relative-error-} that the relative
error at each parameter grid is always smaller 0.2. Therefore, for
normal integrands (not oscillating rapidly), our code is able to yield
acceptable results.
\begin{center}
\begin{table}[H]
\centering{}\caption{\label{tab:Performance-three nodes}Performance for integrands on
three nodes. One node contains 2 K80m, one contains 4 K40m and another
contains one V100. $M=4$, $N=6$ stand for the number of parameters
and the dimension of the integration. The integration is performed
via the direct Monte Carlo method with $10^{5}$ samples. Number of
batches is set to 100. HToD, DToH and GPU evaluation time are averaged
over 100 batches with each batch being evaluated on one GPU. Total,
allocation and retrieve time are averaged over the 10 independent
evaluations.}
{\scriptsize{}}%
\begin{tabular}{cccc}
\hline
 & {\scriptsize{}HToD (ms)} & {\scriptsize{}DToH (ms)} & {\scriptsize{}GPU evaluation time (s)}\tabularnewline
\hline
\hline
{\scriptsize{}K40m (per batch)} & {\scriptsize{}1.42190} & {\scriptsize{}1.71204} & {\scriptsize{}79.62744}\tabularnewline
{\scriptsize{}K80m (per batch)} & {\scriptsize{}1.40043} & {\scriptsize{}3.60045} & {\scriptsize{}89.30734}\tabularnewline
{\scriptsize{}V100 (per batch)} & {\scriptsize{}1.74385} & {\scriptsize{}4.21762} & {\scriptsize{}18.19497}\tabularnewline
\hline
 & {\scriptsize{}allocation time (s)} & {\scriptsize{}retrieve time(s)} & {\scriptsize{}total evaluation time (s)}\tabularnewline
\hline
{\scriptsize{}all three nodes} & {\scriptsize{}0.0415871} & {\scriptsize{}10.6061} & {\scriptsize{}721.20234}\tabularnewline
\hline
\end{tabular}
\end{table}
\par\end{center}

\begin{center}
\begin{figure}[H]
\begin{centering}
\includegraphics[scale=0.45]{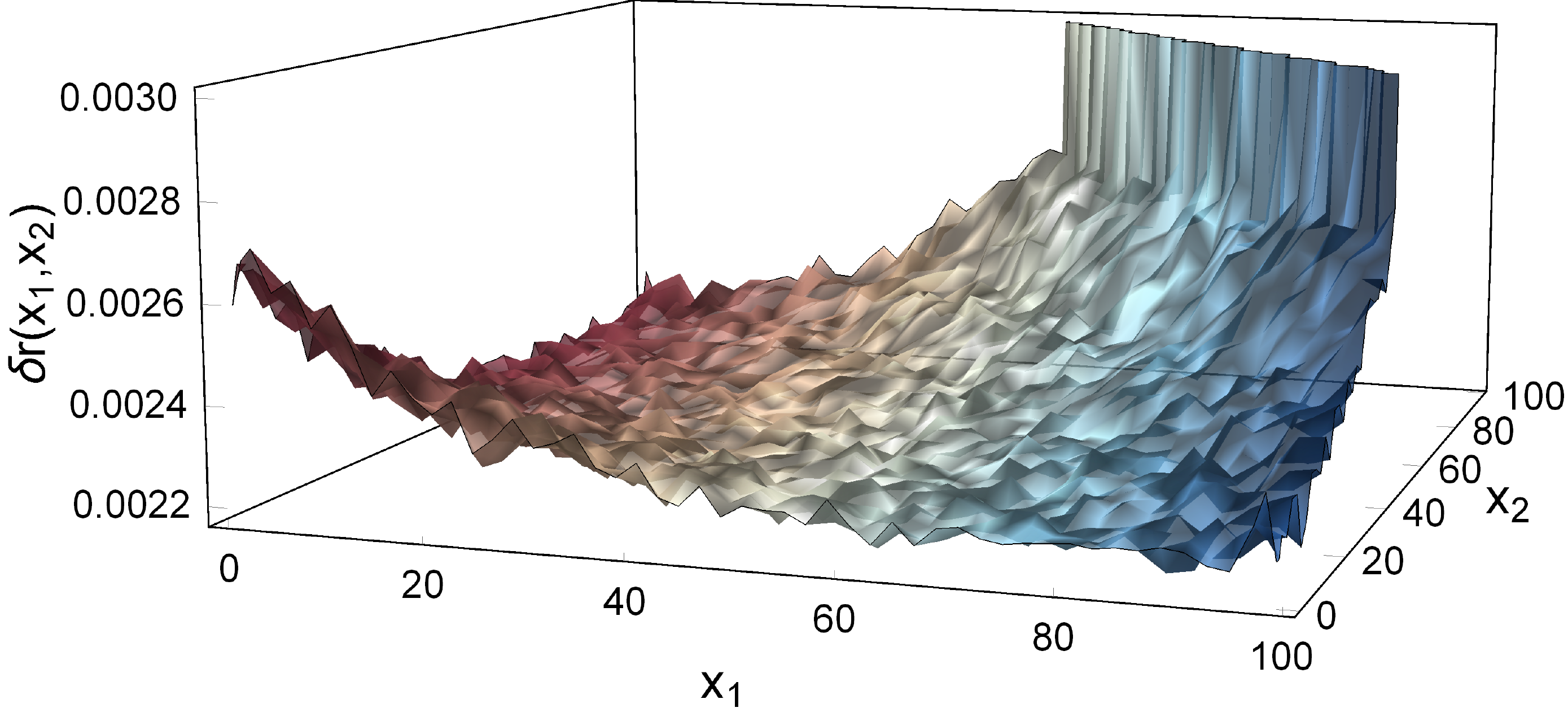}
\par\end{centering}
\centering{}\includegraphics[scale=0.45]{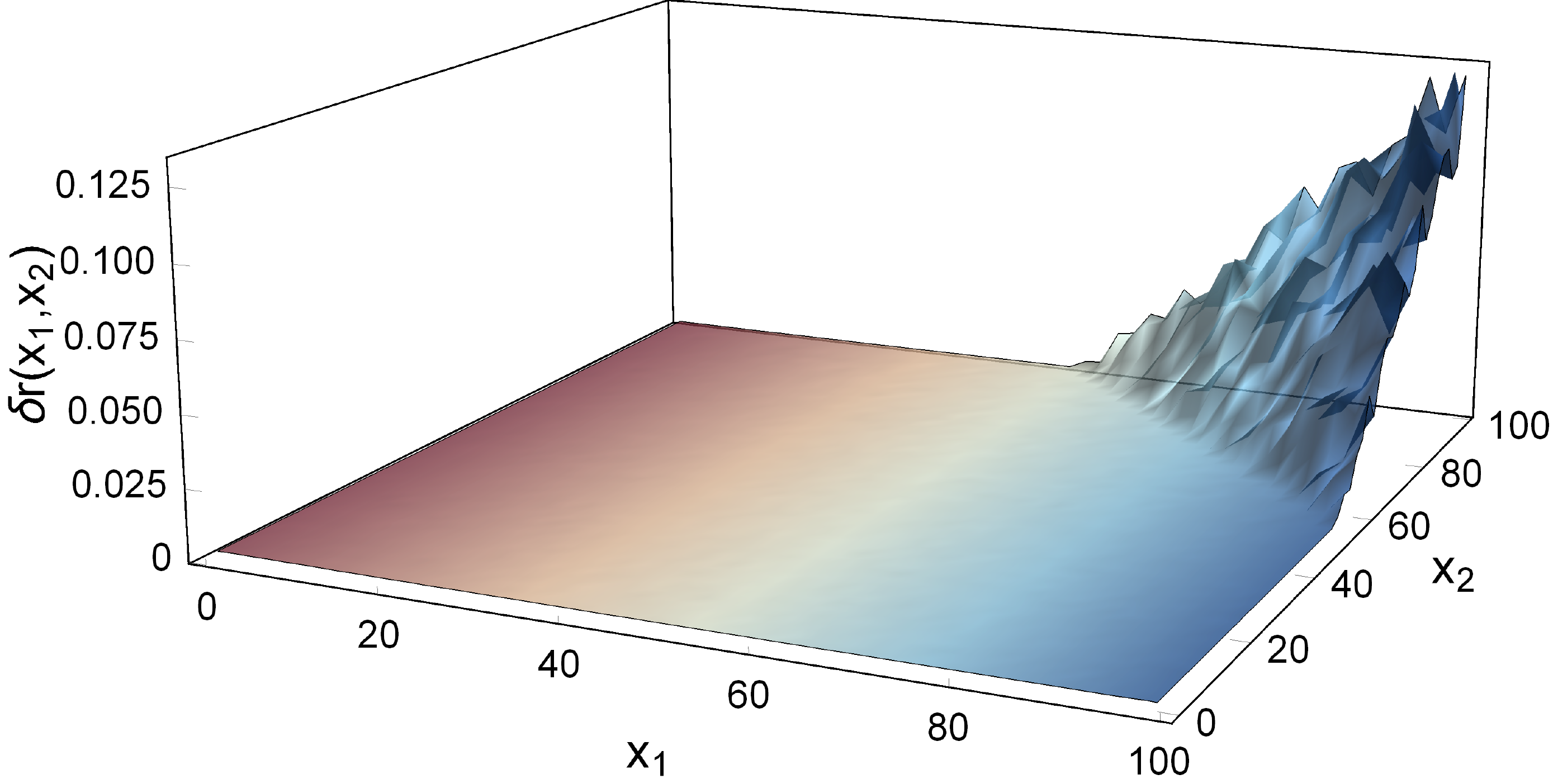}\caption{\label{fig:Relative-error-}Relative error $\delta r(x_{1},x_{2})$
with $x_{1}=x_{2}=\{0,1,2,\cdots99\}$. The cluster contains three
nodes. One node contains 2 K80m, one contains 4 K40m and another contains
one V100. $x_{1}$ and $x_{2}$ expand a parameter grid, and at each
grid value we plot the relative error of the integration. In the upper
panel we have clipped the $\delta r$ axis such that the variations
for low error regions can be seen clearly, while in the lower panel,
we have plotted the full range of $\delta r$.}
\end{figure}
\par\end{center}

\subsection{Functional integration for relativistic Boltzmann equation}

A straight forward application of ZMCintegral-v5 is the solving of
the relativistic Boltzman Equation for Quark Gluon plamsa\cite{ZBEsolver}.
There, we have seven coupled first order differential equations with
an integration of 5 dimensions
\begin{eqnarray}
\frac{\partial\widetilde{f}_{\mathbf{p}}^{a}(x)}{\partial t}+\frac{\mathbf{p}}{E_{\mathbf{p}}^{a}(x)}\cdot\nabla_{x}\widetilde{f}_{\mathbf{p}}^{a}(x)\nonumber \\
-\nabla_{x}\left[E_{\mathbf{p}}^{a}(x)\right]\cdot\nabla_{\mathbf{p}}\widetilde{f}_{\mathbf{p}}^{a}(x) & = & C_{a}(\mathbf{x},\mathbf{p}),\label{eq:boltzmann equation general form}
\end{eqnarray}
where $\widetilde{f}_{\mathbf{p}}^{a}(x)$, the color and spin averaged
distribution function for particle $a$ ($a$ denotes u,d,s,$\bar{\text{u}}$,$\bar{\text{d}}$,$\bar{\text{s}}$
and gluon), is a function of space-time $x^{\mu}=(t,\mathbf{x})$
and momentum $p^{\mu}=(E_{\mathbf{p}},\mathbf{p})$, $C_{a}(\mathbf{x},\mathbf{p})$
is the collision term (5 dimensional integration) for quarks or gluon,
$E_{\mathbf{p}}(\mathbf{x})=\sqrt{\mathbf{p}^{2}+m_{a}^{2}(\mathbf{x})}$.
The complexity of this equation lies in the collision term $C_{a}(\mathbf{x},\mathbf{p})$
which has a large parameter grid.

For Eq. (\ref{eq:boltzmann equation general form}), the parameter
grid is $\widetilde{f}_{\mathbf{p}}^{a}(\mathbf{x})$ where $\mathbf{p}=[p_{x},p_{y},p_{z}]$,
$\mathbf{x}=[x,y,z]$ and $a$ being u,d,s,$\bar{\text{u}}$,$\bar{\text{d}}$,$\bar{\text{s}}$
and gluon. This is a large parameter grid with $[n_{p_{x}},n_{p_{y}},n_{p_{z}},n_{x},n_{y},n_{z}]=[30,30,30,10,10,10]$.
Threfore we have $30^{3}\times10^{3}\times7\sim2\times10^{8}$ grid
points to scan. When evaluate $C_{a}(\mathbf{x},\mathbf{p})$, which
is a 5 dimensional integration, since $\mathbf{x}$ and $\mathbf{p}$
are parameters, we need to evaluate $2\times10^{8}$ $C_{a}$. For
this specific task, our previous versions perform poorly since users
have to provide the parameter grid in CPU, and evaluate each integration
in GPU one by one. The newest version, which returns the entire parameter
grid from GPU with each element being the integrated values, saves
much of the communication time between Host (CPU) and Device (GPU).
Thus it is more suitable for large parameter scan and relatively lower
dimensional integrations.

\section{Conclusion and Discussion}

To meet the requirement of integration with large parameter grids,
we have added a new functionality to ZMCintegral-5 which is able to
give the integration results at each grid point on multi-GPUs. The
code supports user defined functions and is easy to use in the Python
language. To ensure the calculation speed, we only adopt the direct
Monte Carlo method to evaluate integrations. For the parameter scan
functionality, we suggest a number of sample points not exceed $10^{6}$
(for parameter grid size \textasciitilde{} $10^{8}$) for one Tesla
V100. However, for users with large GPU clusters, this number can
be set higher. The time consumption for data transfer from host to
device and between different nodes is negligible compared with the
GPU evaluation time. Therefore, users only need to consider the number
of sample points, which determines the accuracy and calculation time
of the integration task.

\section*{\textit{Acknowledgment}.}

The authors are supported in part by the Major State Basic Research
Development Program (973 Program) in China under Grant No. 2015CB856902
and by the National Natural Science Foundation of China (NSFC) under
Grant No. 11535012. The Computations are performed at the GPU servers
of department of modern physics at USTC. We are thankful for the valuable
discussions with Prof. Qun Wang of department of modern physics at
USTC.

\end{small}


\label{}



\label{-1}





\bibliographystyle{elsarticle-num}
\nocite{*}
\bibliography{ZMCintegral}






\end{document}